\documentclass[aps,prl,reprint,groupedaddress]{revtex4-1}

\usepackage{graphicx}
\usepackage{color}
\begin{document}

\title{Glassy magnetic behavior in the metamagnetic DyAlO$_3$ doped with Cr}

\author{R. Escudero, B. L. Ruiz-Herrera, M. P. Jimenez,  and F. Morales}
\email[Author to whom correspondence should be addressed. FM, email address: ] {fmleal@unam.mx}
\affiliation{Instituto de Investigaciones en Materiales, Universidad Nacional Aut\'{o}noma de M\'{e}xico. M\'{e}xico, D.F., 04510 M\'EXICO.}


\begin{abstract}
Magnetic properties of DyAl$_{0.926}$Cr$_{0.074}$O$_3$ and DyAlO$_3$ were studied. We found that both compounds are antiferromagnetic with a low N\'eel transition temperature. At higher temperatures the magnetic characteristics show a Curie-Weiss dependence.
The N\'eel temperature disappears when a field of about 2 T is applied, the system changes from an antiferromagnetic to a weak ferromagnetic behavior due to a metamagnetic transition.  Furthermore, AC magnetic measurements in the Cr doped compound, at different frequencies, show a spin glass-like behavior. These transitions were studied and corroborated by specific heat measurements. We found the presence of metamagnetism and spin glass in the compound doped with chromium, determining that the small addition of chromium atoms modifies the magnetic properties of the compound DyAlO$_3$, resulting in new features such as the spin glass-like behavior.
\end{abstract}

\pacs{Perovskite, Magnetic properties, Spin glass, Metamagnetism}

\maketitle

\section{Introduction}
The large family of compounds known as perovskites has a simple crystal structure, related to the mineral CaTiO$_3$. This family forms an enormous variety of compounds with a very rich variant in its chemical and physical properties. The base compound, ABX$_3$ type, presents a cubic crystal structure in his ideal form. This can be described as corner sharing BX$_6$ octahedral where the A cation, that is in the middle of the structure, has 12-fold coordination site \cite{johnns,balla}.

Perovskites show many structural modifications that according to the used atoms give place to a great variety of magnetic and electronic characteristics. An example of a compound whose structure present a distortion of the ideal perovskite form, is the orthorhombic DyAlO$_3$ \cite{vasylechko,dalziel60}, which magnetic behavior is widely known \cite{petrov11,holmes,holmes72,schu,Cashion68,Bidaux68}, and interpreted from the viewpoint of the Anderson model of magnetic exchange in insulators, with a mechanism of indirect exchange between the nearest Dy ions \cite{petrov12}.

According to  neutron diffraction study of the compound \cite{Bidaux68}, the dysprosium atoms form a magnetic structure of type $G_xA_y$, in Bertaut's notation, which corresponds to the space group \textit{Pbnm}. Magnetic order generated by dysprosium resembles two antiferromagnetic networks embedded, one of them canted relative to one another. DyAlO$_3$ has this antiferromagnetic behavior below a $T_N\sim$3.5 K. Moreover DyAlO$_3$ exhibits a metamagnetic transition that is related to an abrupt change of one or more magnetic sublattices under an applied field \cite{petrov11,holmes72,schu}.

The subclass of magnetic materials known as the spin glass materials, consist in a mixed-interacting magnetic system formed by magnetic moments randomly located in a lattice, leading to multidegenerate ground states but also a cooperative freezing transition at a well defined temperature, the freezing temperature $T_f$ \cite{mydosh93,rmp,blundell}.

At high temperature all magnetic moments in a spin glass are independent. As the temperature decreases some spins build up into locally correlated units, spins that are not included in this formed clusters take part interacting between them.  Finally at $T_f$ the system achieve one of its many ground states and freezes, process that is understood as a cooperative effect. Consequently below $T_f$ a spin glass posses metastability reflecting in a divergence between the field-cooled and zero field-cooled magnetic susceptibility below $T_f$, among other features.

 Perovskites of the type TRMO$_3$ (TR = rare earth, M = transition metal) with partial substitution of the transition metal or the rare earth, show spin glass behavior \cite{perez98,nam00,motohashi05}. DyAlO$_3$ is an antiferromagnetic material and its magnetic behavior with partial substitutions has not been determined yet.

In this paper we report magnetic measurements on the orthorhombic  perovskites; DyAl$_{0.926}$Cr$_{0.074}$O$_3$ and DyAlO$_3$. The small amount of Cr was eventually used to observe the behavior of the compound, and also in other  lanthanide perovskites as magnetic pigments, however we must stress that the spin glass behavior was a serendipitous discover, and was only observed in this compound, more different compositions will be studied in the future. Magnetic properties were carried out by DC and AC molar susceptibility measurements. The DC susceptibility was measured at different magnetic fields and temperatures, meanwhile the AC susceptibility was performed as function of temperature and frequency. In addition, specific heat measurements as function of temperature and magnetic field were performed. The main results of this report are that in the doped compound the metamagnetism persist and a spin glass behavior is developed below 3.2 K. We analyze the spin glass$-$like behavior using the conventional critical slowing down spin dynamics.

\begin{table*}
\caption{\label{table1}Crystallographic data for DyAl$_{0.926}$Cr$_{0.074}$O$_3$ and DyAlO$_3$ obtained by Rietveld refinement. Symmetry is described by the orthorhombic space group $Pbnm$,  standard deviations are written between parentheses.}
\begin{tabular}{c c c c c }
\hline
   &  & {\bf DyAl$_{0.926}$Cr$_{0.074}$O$_3$ }&  &  \\
\hline
Rp (\%) & Rwp (\%) & Re (\%) & $\chi ^2$ &  \\
20.7 & 29.8 & 24.81 & 1.44 &  \\
\hline
Parameters (\AA) & $a$ = 5.33423(3) & $b$ = 7.40830(4) & $c$ = 5.21155(3) &  \\
Volume (\AA$^3$) & 205.948(0.002) &  &  &  \\
\hline
Site & \textit{x} & \textit{y} & \textit{z} & Occupation \\
 Al & 0 & 0 & 0 & 0.463(3) \\
 Dy & 0.44982(11) & 0.25000(0) & -0.00999(21) & 0.500(0) \\
 O1 & 0.02531(121) & 0.25000(0) & 0.07013(136) & 0.500(0) \\
 O2 & 0.29257(108) & -0.04074( 76) & 0.21421(115)   & 1.000(0) \\
 Cr & 0 & 0 & 0 & 0.037(3) \\
\hline
   & DyAl$_{0.926}$Cr$_{0.074}$O$_3$ & Dy$_2$O$_3$ &  & \\
Phase fraction (\%) & 98.10(0.29) & 1.90(0.04) &         & \\
\hline
  &  & {\bf DyAlO$_3$ }&  &  \\
  \hline
  Rp (\%) & Rwp (\%) & Re (\%) & $\chi ^2$ &  \\
 22.0 & 29.8 & 23.28 & 1.63 &  \\
\hline
 Parameters (\AA) & $a$ = 5.31986(3) & $b$ = 7.40187(5) & $c$ = 5.21007(3) &  \\
 Volume (\AA$^3$) & 205.157(0.002) &  &  &  \\
\hline
 Site & \textit{x} & \textit{y} & \textit{z} & Occupation \\
 Al & 0 & 0 & 0 & 0.500(0) \\
 Dy & 0.45209(11) & 0.25000(0) & -0.00953(20) & 0.500(0) \\
 O1 & 0.00769(112) & 0.25000(0) & 0.06031(134) & 0.500(0) \\
 O2 & 0.28016(115) & -0.03580(74) & 0.21425(106) & 1.000(0) \\
 \hline
   & DyAlO$_3$ & Dy$_2$O$_3$ &  & \\
 Phase fraction (\%) & 93.27(0.26) & 6.73(0.06) & & \\
\end{tabular}
\end{table*}

\section{Experimental details}

The materials used for the synthesis were: Dy$_2$O$_3$, Al(NO$_3$)$_3\cdot$ 9H$_2$O and Cr$_2$O$_3$ provided by  J. T. Baker. Compounds were prepared by mixing the starting materials in the relations to the chemical formula: DyAl$_{0.93}$Cr$_{0.07}$O$_3$, and DyAlO$_3$. All powders were mixed in an agate mortar. Platinum crucibles were used. Residual compounds as water and nitrogen oxides were eliminated by the thermal treatment. Samples were annealed in a furnace at 900$^{\circ}$ C in a period of 24 h, after this first treatment the resulting powders were fired at temperatures about 1000$^{\circ}$ C in a period of 96 h. Samples were characterized by X-ray powder diffraction (XRD) with a Siemens D5000 diffractometer, using Co $K_{\alpha 1}$ wavelength and a secondary graphite monochromator. Measurements were carried out on the 2$\theta$ range from 5 to 110 degrees. Rietveld analysis \cite{rietveld} was performed, using the \textit{Fullprof} software, in order to distinguish more clearly the influence of chromium atoms and characterization of all crystal cell parameters.

Magnetic measurements were performed using a MPMS (Magnetic Properties Measurements System) Quantum Design magnetometer. The range of measured temperature was from 2 K to 300 K and the applied magnetic field was about $\pm$ 5 T. Measurements were performed in zero field cooling (ZFC) and field cooling (FC) modes. Isothermal magnetic measurements were performed between $\pm$ 5 T, for sample doped with Cr at 2 K and 3.2 K and for DyAlO$_3$ at 2 K and 5 K. AC magnetic measurements were performed, in both compounds, with an applied magnetic field of 1 Oe and frequencies from 50 Hz to 1000 Hz. Furthermore, specific heat $C_P$ measurements were performed in a PPMS of Quantum Design at zero magnetic field and high magnetic fields up to 9 T. In these measurements the specific heat addenda, sample support and grease, was extracted to obtain the absolute $C_P$ of the sample.

\begin{figure}[btp]
\includegraphics[scale=0.3]{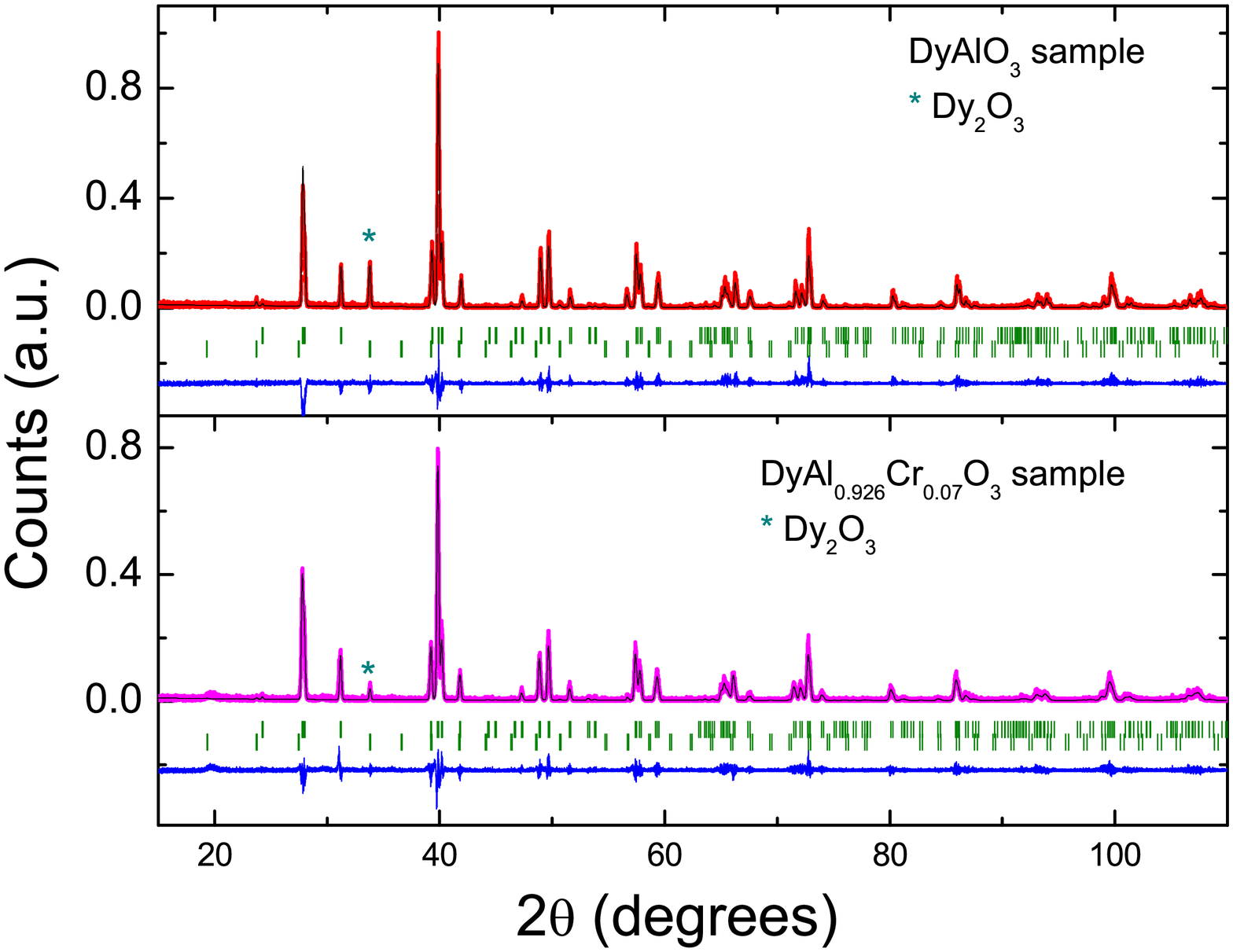}
\caption{(Color online) Rietveld refinement of the DyAlO$_3$ and DyAl$_{0.926}$Cr$_{0.074}$O$_3$ samples. Red and purple lines  are the experimental data, the superposed black line is the calculated pattern. At the bottom of each diagram, the difference  between  experimental and calculated data is shown in blue line. Vertical green marks are displayed in two rows, upper ones correspond to the Bragg positions for DyAlO$_3$ and lower ones to Dy$_2$O$_3$ patterns.}
\label{refinado}
\end{figure}

\section{Results and discussion}

\subsection{Structural characterization}

The examination of X-ray diffractograms shows that the perovskites based structures DyAl$_{0.93}$Cr$_{0.07}$O$_3$, and DyAlO$_3$ have orthorhombic symmetry with $Pbnm$ space group. Figure \ref{refinado} shows the results of the Rietveld refinement and table \ref{table1} contains the crystallographic data of the samples.  The occupation factor reveals that the chromium resides in the same position of aluminium atoms in the unit cell, besides, give us information about the quantity of Al ions that are being substituted by  Cr. Accordingly, the stoichiometry changes gave the resulting formula DyAl$_{0.926}$Cr$_{0.074}$O$_3$. This will be used as the real stoichiometry in all the text. The polycrystalline phases were identified by comparison with X-ray patterns in the Inorganic Crystal Structure Database (ICSD) 2012. All peaks correspond to the DyAlO$_3$ phase \cite{dalziel60}, except the peaks related to Dy$_2$O$_3$ found in low proportion in the compound, however according to our results this impurity is not affecting the observed magnetic characteristics. Rietveld refinement was made using crystallographic data of the isostructural compound GdAlO$_3$(ICSD 59848) and Dy$_2$O$_3$ (ICSD 82421). In the distorted perovskites Al$^{3+}$ and Cr$^{3+}$ ions remain essentially in octahedral sites; thus when Al$^{3+}$ is substituted by Cr$^{3+}$, which is about 12.3\% bigger than the size of Al$^{3+}$, the Al-O distance decreases. Despite the inclusion of chromium, the volume of unit cell does not present a considerable change, only about 0.4 \%.

\begin{figure}[btp]
\begin{center}
\includegraphics[scale=0.3]{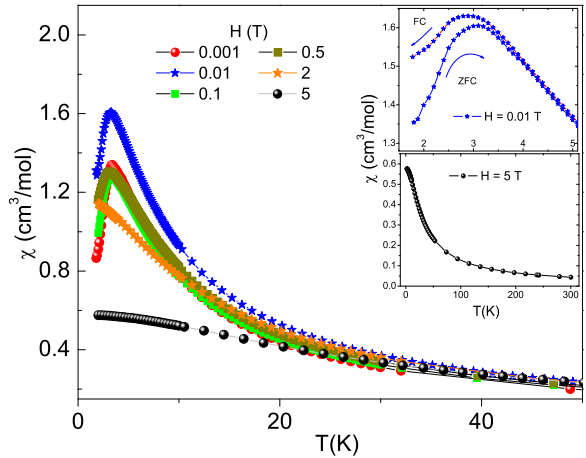}
\end{center}
\caption{(Color online) $\chi(T)$ in ZFC mode for the compound DyAl$_{0.926}$Cr$_{0.074}$O$_3$ determined at several magnetic fields from 0.001 to 5 T. The main panel shows $\chi(T)$ from 2 to 50 K. The peak observed at low field indicates the antiferromagnetic transition at about 3.2 K. The upper inset displays the irreversible behavior of the curve with measurements performed in FC and ZFC modes at 0.01 T. Lower inset shows the behavior at H = 5 T in ZFC mode from 2 K to 300 K.}
\label{XT_DC}
\end{figure}

\subsection{Magnetic measurements}

In Fig. \ref{XT_DC} the main panel displays the magnetic susceptibility, $\chi(T)$, for the compound DyAl$_{0.926}$Cr$_{0.074}$O$_3$, from 50 to 2 K at six different magnetic fields. It could be thought that the peak at about 3.2 K is only the signature of an antiferromagnetic transition, $T_N$, due that the undoped compound has a similar transition at 3.55 K corresponding to the N\'eel temperature \cite{Bidaux68,schu}. However, a careful observation shows that the peak is quite broad, and also the size of the peak is very dependent of the intensity of the applied magnetic field: it is smoothed with increasing field and finally at 2 T disappears. Thus, accordingly to this, the peak may be related to an antiferromagnetic transition with another physical process associated.

The upper inset in Fig. \ref{XT_DC} shows measurements in ZFC and FC modes at low temperature. The different behavior of the ZFC and FC curves, besides the width of the peak, suggest us a spin glass behavior. Careful examination of  $\chi(T)$ measurements at 5 T shows a negative curvature at low temperature, clearly seen in the main panel and in the lower inset of Fig. \ref{XT_DC}. This change must be compared with the curves measured at 0.001, 0.01, 0.1, and 0.5 T. The change of curvature is also an indication of a metamagnetic transition induced by the magnetic field; thus, a change from canted antiferromagnetism to a weak ferromagnetism.

\begin{figure}[btp]
\begin{center}
\includegraphics[scale=0.3]{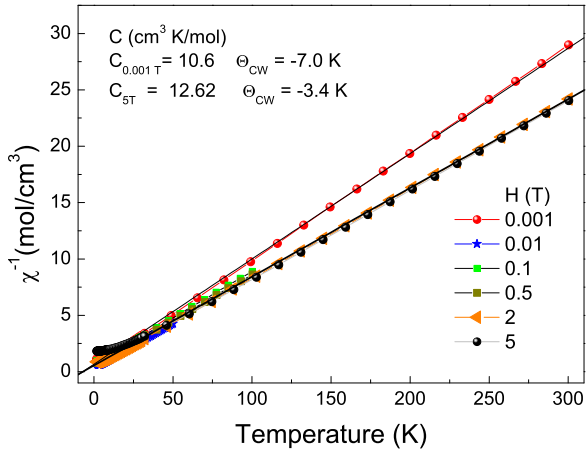}
\end{center}
\caption{(Color online) Inverse of the magnetic susceptibility, $\chi(T)^{-1}$ measured at different magnetic fields for DyAl$_{0.926}$Cr$_{0.074}$O$_3$. Lines on the experimental data are the fitting to the Curie-Weiss law at two fields, 0.0001 and 5 T. C`s are the Curie constants, and Curie-Weiss temperatures are the resulting values of the fitting.}
\label{Xinv_DC}
\end{figure}

\begin{figure}[btp]
\begin{center}
\includegraphics[scale=0.3]{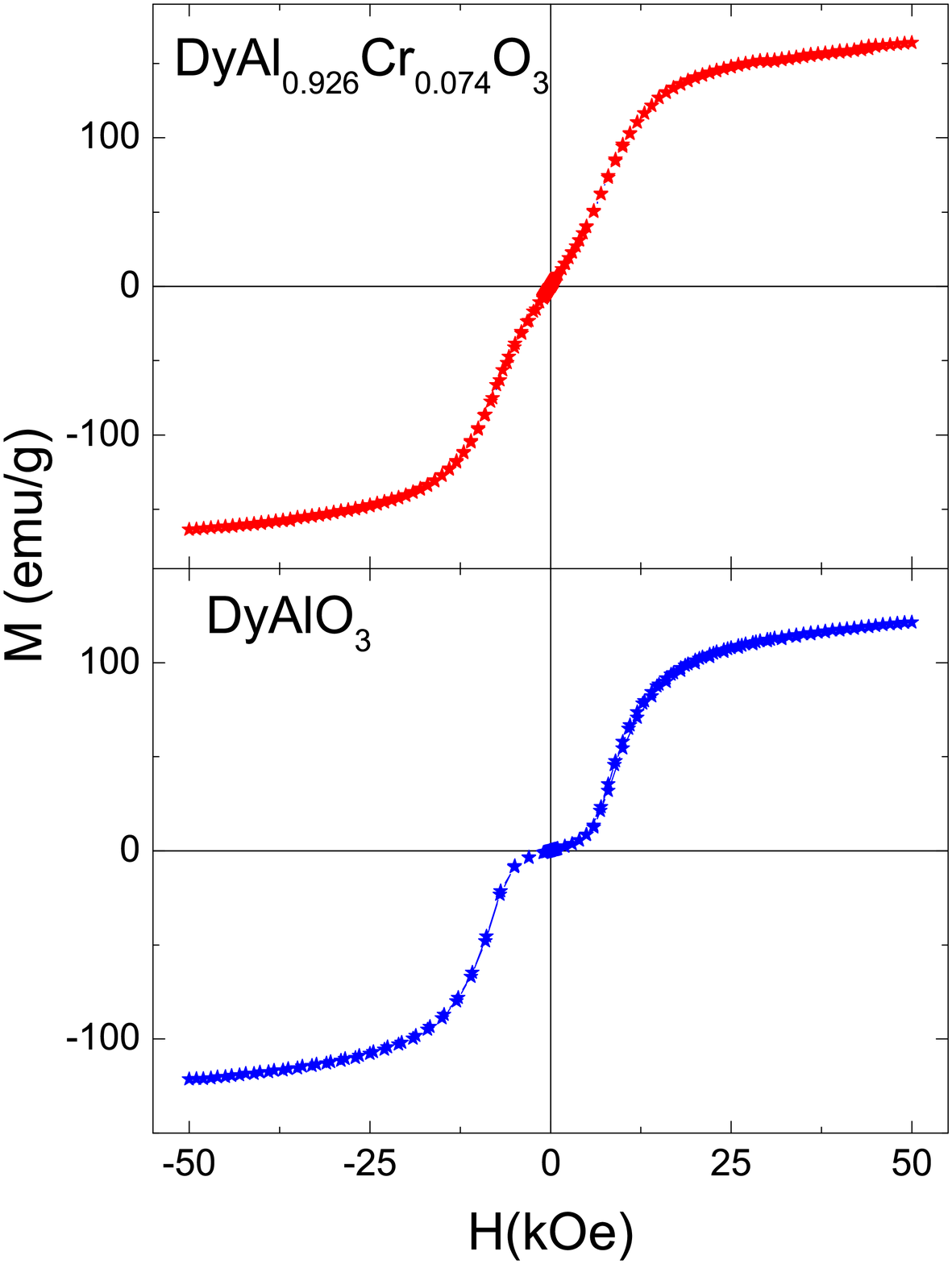}
\end{center}
\caption{(Color online) The top panel presents the metamagnetic change of the doped Cr compound in $M(H)$, whereas the lower panel displays the well known metamagnetic transition of DyAlO$_3$. Data of two panels were taken at 2 K below the N\'eel Temperature. It is interesting to compare both figures, in the Cr doped sample the metamagnetic behavior is clearly seen but quite reduced as compared with the pure compound.}
\label{MH_Cr}
\end{figure}

The inverse of the susceptibility $\chi(T)^{-1}$ of DyAl$_{0.926}$Cr$_{0.074}$O$_3$, provides more information about the magnetic behavior. The experimental data above 100 K were fitted (continuous line) with the Curie-Weiss law, as illustrated in Fig. \ref{Xinv_DC}. From the fitting we obtain the Curie constant, $C$, and the Curie-Weiss temperature, $\theta_{CW}$. Measurements of the specimen in 0.001 T give $C$=10.6 cm$^{3}$ K mol$^{-1}$, and $\theta_{CW}$=-7 K;  whereas, the measurement performed at 5 T gives $C$=12.62 cm$^{3}$ K mol$^{-1}$, and $\theta_{CW}$=-3.4 K. Those changes in the constants can be correlated to the shift of the magnetic behavior in a very clear manner. Magnetic effective  moments, $\mu_{eff}$, Curie constants, and Curie-Weiss temperatures are slightly dependent on the applied magnetic field, showing only very small changes: for low field, 0.001 T, $\mu_{eff} = 9.6$ $\mu _B$, whereas for 5 T $\mu_{eff}=10.03$ $\mu _B$. These values are similar to the theoretical value of Dy$^{3+}$ of 10.65 $\mu_B$. Experimental values of $\mu_{eff}$ of DyAlO$_3$ reported are 9.2 $\mu_B $ \cite{holmes} and 0.38 $\mu_B $ in nanocrystals \cite{petrov11}.

 \begin{figure}[b]
\begin{center}
\includegraphics[scale=0.3]{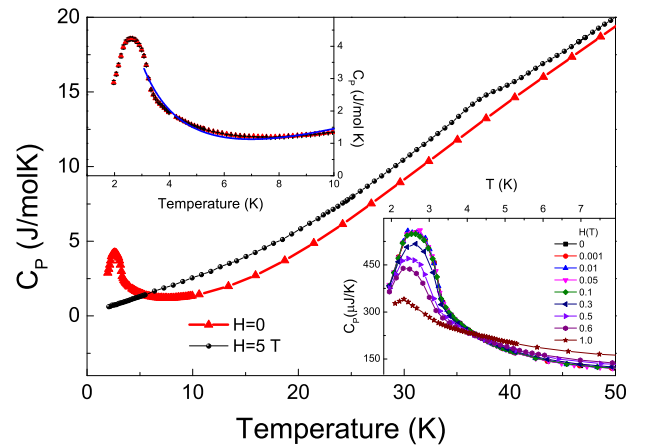}
\end{center}
\caption{(Color online) The main panel displays the specific heat of DyAl$_{0.926}$Cr$_{0.074}$O$_3$ from 2-50 K at two magnetic fields: 0, and 5 T. Note the suppression of the N\'eel peak at low temperature, but also a bump at a temperature about 38 K at H = 5 T. Top insert shows a fitting (continuos line) of the data at low temperatures. The lower insert displays a close view of the influence of the magnetic field and the suppression of the peak at 2.3 K. }
\label{Cp}
\end{figure}

 Figure \ref{MH_Cr} shows the magnetization as a function of magnetic applied field $M(H)$ at 2 K, below the N\'eel temperature. When the field is increased and decreased for the doped and pure compounds, shown in the top and low panel respectively, a slope change is observed about $\pm$10 kOe, this change corresponds to the metamagnetic transition. This transition agrees well to previously reported one for DyAlO$_3$ \cite{petrov11}.  We must stress the persistence of the metamagnetism in the Cr doped compound, although it is quite reduced. The reduced metamagnetism in the doped compound compared to the undoped one may be explained by the distortions created by the Cr magnetic moment. The small amount of Cr, introduces small distortions in the magnetic structure in  the three magnetic subcells of the compound, but without completely destroying the metamagnetic behavior, The Cr introduces disorder that modify the magnetic interactions between Dy atoms. Moreover, it is shown that saturation magnetization of compound doped with Cr is higher than non doped one.

\subsection{Specific heat and AC susceptibility}

Specific heat measurements provide information about magnetic transitions. The specific heat of DyAlO$_3$ has a transition lambda type at about 3.5 K \cite{Tanaeva03,Cashion68}, in agreement to the magnetic measurements. The transition observed in DyAl$_{0.926}$Cr$_{0.074}$O$_3$ (Fig. \ref{Cp}) is not of the lambda type, suggesting that Cr produces some effect on the magnetic order. Fig. \ref{Cp} presents  specific heat measurements at two magnetic fields; 0 and 5 T. The main panel displays these measurements from 2 - 50 K. The red curve (triangles) shows the measurements at zero magnetic field, whereas the black dots are the measurements with  field at 5 T. Two important changes must be noted: at zero field and low temperature a peak is clearly seen and is associated with the antiferromagnetic transition, this peak was also seen  in the $\chi(T)$ measurements. However, with 5 T the peak at low temperature disappears, and a small bump is present at about 38 - 40 K.

Assuming that DyAl$_{0.926}$Cr$_{0.074}$O$_3$ is an electric insulator the specific heat data were fitted between 3.1 K and 10 K with the following equation \cite{tari}:

\[C_P=\alpha T^{-2}+ \beta_3T^3 + \beta_5T^5 + \beta_7T^7,
\]
the first term represents the contribution to $C_P$ because of the short-range-order effect of the spin alignment when $T$ approach to $T_N$ from high temperature \cite{miyazaki05}. The second term is the harmonic contribution of the lattice vibrations, whereas the last two terms are due to quasi-harmonic contributions to the specific heat\cite{dug}. Those terms give $\beta$ values very small but important in the case of anharmonicity produced by impurities or disorder as we considered for the case produced by the chromium atoms.  The upper inset in Fig. \ref{Cp} displays the  curve portion of $C_P-T$ data and the obtained fit, continuous line. From the fitting we obtain; $\alpha=(30.88 \pm 0.374$) J K/mol; $\beta_3=(0.00181 \pm 8.8\times 10^{-5}$) J/mol K$^4$; $\beta_5=(-7.86\times 10^{-6} \pm 8.6\times 10^{-7}$) J/mol K$^6$; $\beta_7=(1.23\times 10^{-8} \pm 1.86\times 10^{-9}$) J/mol K$^8$. Using the Debye approximation; $\beta_3=1973.7 s\theta_D^{-3}$, where $s$ is the number of atoms per formula unit, we obtain $\theta_D = 175$ K, in agreement to $\theta_D$ values reported for DyAlO$_3$ \cite{cashion98,kuzmin93}. The lower inset of Fig. 5 displays the magnetic effect on the anomalous peak associated to the N\'eel temperature. Note the variation of the size of the peak as the field is increased, this peak is quite width and behaves differently to normal antiferromagnetic peak. In fact this anomalous shape take us to perform AC magnetization measurements  as a function of frequency, study that is showed in the last part of this section.

\begin{figure}[bth]
\begin{center}
\includegraphics[scale=0.3]{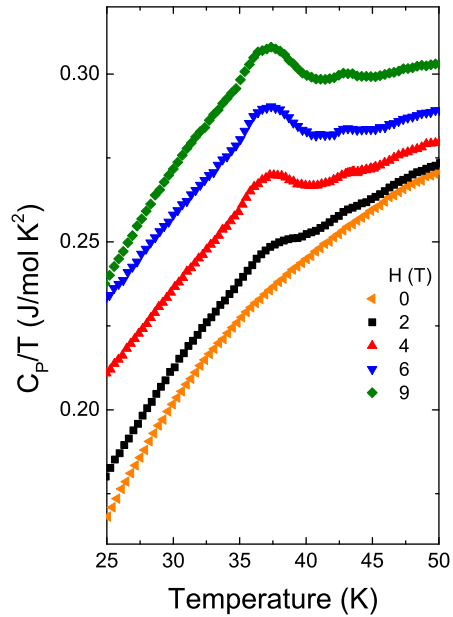}
\end{center}
\caption{(Color online) Plot of $C_P/T-T$ from 50 to  25 K of DyAl$_{0.926}$Cr$_{0.074}$O$_3$, measured at different magnetic fields, from 0 up to 9 T. It is clearly seen a bump arising at a temperature about 37.3 K and a smoother one around 43 K.}
\label{CpVsT}
\end{figure}

Fig. \ref{CpVsT} shows the influence of the magnetic field in specific heat measurements at temperatures  between 50 K and 25 K, in this figure we plotted C$_P$/T versus T. In this temperature range two bumps arise with a field of 2 T, at temperature about 37.3 K and 43.4 K, increasing in size as the magnetic field was increased. It is necessary to remark that in the curve measured at zero field there is no feature at all.
We must also mention that a Schottky anomaly reported in Dy$_2$O$_3$, at 45 K could be related to this feature  \cite{gruber82}, however it is also clear that if the Schottky anomaly has some influence this necessary will be present without magnetic field.
Because of the magnetic measurements do not show any one feature, at around the temperatures where the bumps are observed in $C_P(T)$, we think that  these bumps could be related to a structural modification produced by the magnetic field, as was observed on single crystals of DyCrO3 \cite{Krynetski97}.

\begin{figure}[b]
\begin{center}
\includegraphics[scale=0.4]{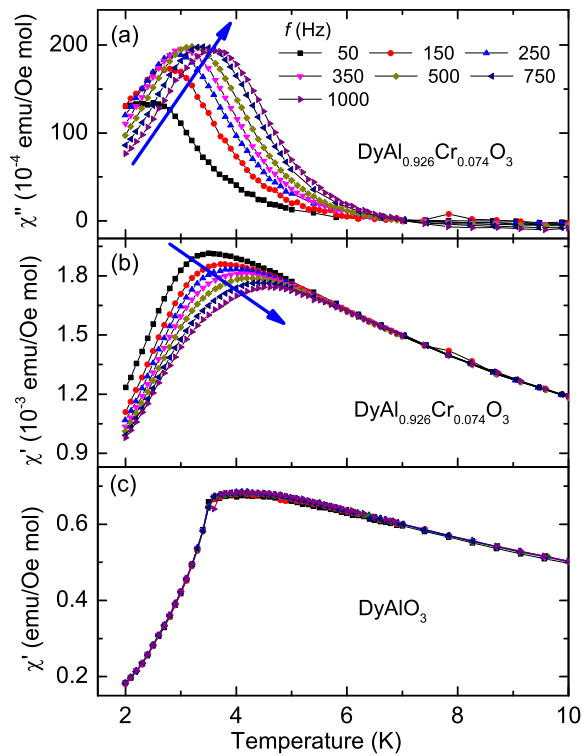}
\end{center}
\caption{(Color online)  Imaginary part (a) and the real part (b) of the AC susceptibility  of DyAl$_{0.926}$Cr$_{0.074}$O$_3$, at different frequencies showing spin glass characteristics. (c) shows the real part of the AC susceptibility $\chi'(T)$ of DyAlO$_3$.}
\label{XT_AC}
\end{figure}

 Fig. \ref{XT_AC}(a) and (b) displays the real part $\chi'$ and imaginary part $\chi''$ of the AC susceptibility as a function of temperature and frequency, $f$, of the doped compound. The measurements were performed at seven different frequencies. The maximum of $\chi'(T)$ displaces to higher temperatures as the frequency increases. This behavior has been associated to spin glass systems. It is noteworthy that above 6 K, $\chi'(T)$ becomes frequency independent. Figure \ref{XT_AC}(c) displays $\chi'(T)$ of DyAlO$_3$, its behavior belongs to an antiferromagnetic material. There is not a frequency effect on the maxima of the AC susceptibility,  demonstrating that the insertion of Cr in the lattice provokes spin glass behavior. It is noteworthy to mention that there are few previous studies  where spin glass, antiferromagnetism and metamagnetism are present in the same specimen.

\begin{figure}[btp]
\begin{center}
\includegraphics[scale=0.3]{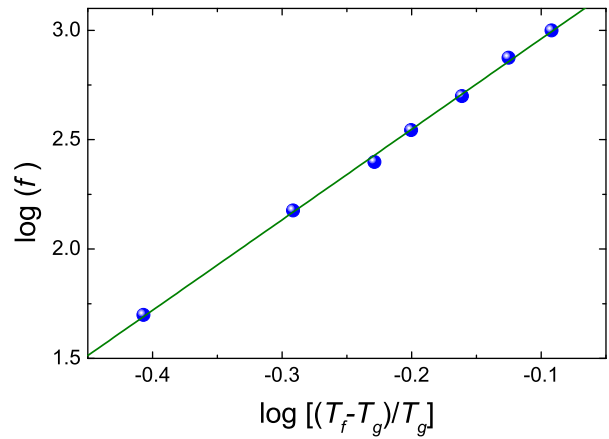}
\end{center}
\caption{(Color online) Dynamic scaling of the critical slowing down  of the compound DyAl$_{0.926}$Cr$_{0.074}$O$_3$ extracted from  $\chi'(T)$ data at various frequencies $f$. }
\label{Dynamic}
\end{figure}

AC susceptibility measurements as a function of frequency permit to obtain information about the dynamic behavior of spin glasses using the standard critical slowing-down formula \cite{binder86}:

\[\frac{\tau}{\tau_0} = \left( \frac{T_f - T_g}{T_g}\right)^{-z\nu}, \]
where $\tau=(2\pi f)^{-1}$ is the relaxation time, $z\nu$ is a  critical exponent, and $\tau_0$ is a microscopic relaxation time  corresponding to the shorter time available to the fluctuations. $T_f$ is the freezing temperature, defined as the temperature at the maximum of $\chi'(T)$, meanwhile $T_g$ is the critical temperature for spin glass ordering equivalent to $T_{f}(\omega)$ as $\omega \rightarrow 0$ \cite{gunnarsson88,malinowski11}.
The main panel of Fig. \ref{Dynamic} displays log($f$) versus log[($T_f - T_g$)/$T_g$] for the different  frequencies used in the AC susceptibility measurements. The straight line is the best fit to the data  obtained with; $T_g\simeq 2.5$ K, $z\nu\simeq 4.1$ and $\tau_0 \simeq 6.7 \times 10^{-5}$ s. The value of $z\nu$  for several spin glasses is between the values 4 to 12 \cite{imamura08}. However, the values expected for $\tau_0$ in  conventional spin-glass are of the order of $10^{-13}$ s, shorter than the value obtained for  DyAl$_{0.926}$Cr$_{0.074}$O$_3$. $\tau_0$ values of the order of $10^{-10}$ s have been reported in systems where formation of glazing clusters with ferromagnetic order \cite{nam00,tang06}. Values of $\tau_0 \sim 10^{-5}$ and $z\nu\approx1$ have been reported in compounds where two dynamical freezing processes are in the same system \cite{de06}.

DC and AC magnetization measurements on DyAl$_{0.926}$Cr$_{0.074}$O$_3$ at low temperature and low magnetic field suggest that antiferromagnetism and spin-glass coexists below $T_g$. As mentioned before the antiferromagnetic state was produced by the Dy sub-cells and the spin glass is produced by the Al network with Cr impurities. One last point worth to mention is the fact that the DC magnetic measurements at low field, in the Cr doped compound, show $T_N=3.1$ K, whereas the analysis of the relation of frequency and the freezing temperature gives $T_g=2.5$ K that coincides with the maximum of $C_P(T)$ at low field and low temperature.

\section{Conclusions}

Two Dy orthoaluminate perovskites with space group {\sl Pbnm}, formulae DyAlO$_3$ and DyAl$_{0.926}$Cr$_{0.074}$O$_3$, were studied and compared. The Cr doped compound maintains the metamagnetic transition observed in DyAlO$_3$, but quite reduced and changing from an antiferromagnetic behavior to a weak ferromagnetism.  The ac susceptibility measurements indicate an unusual spin glass-like behavior as the dynamic analysis shows; a long relaxation time that indicates the presence of two dynamical freezing processes. This different behavior respect to other spin glasses may be because the complicated magnetic structure of the compound, as already studied by other researchers. Interesting magnetic features were observed in the specific heat measurements at about 35-45 K, with increasing magnetic field. These  magnetic anomalies observed in our specific heat measurements are indicative of the complicated magnetic structure of the compound. All these observations situate the DyAl$_{0.926}$Cr$_{0.074}$O$_3$ as an uncommon  glassy compound that requires further investigations.

\begin{acknowledgements}
Partial support for this work is gratefully acknowledge to CONACyT Project 129293(Ciencia B\'{a}sica), DGAPA-UNAM project No. IN100711, project BISNANO 2011, and project PICCO 11-7 by The Institute of Sciences of Distrito Federal, Ciudad de M\'{e}xico.
We Thank to Mr. R. Reyes, for help in technical assistance, and to Dr. C. Pi\~na for the initial prepared samples.
\end{acknowledgements}


\thebibliography{99}
\bibitem{johnns} Johnsson M, Lemmens P (2007), Crystallography and Chemistry of Perovskites. In: Kronm\"{u}ller H, Parkin S (eds) Handbook of Magnetism and Advanced Magnetic Materials, John Wiley \& Sons Ltd, UK, pp 2098-2106,
\bibitem{balla} Bhalla AS, Guo R, Roy R (2000) The Perovskite Structure - A Review of its Role in Ceramic Science and Technology, Mat Res Innovat 4:3-26
\bibitem{vasylechko} Vasylechko L, Senyshyn A, Bismayer U (2009) Perovskite-Type Aluminates and Gallates. In: Handbook on the Physics and Chemistry of Rare Earths, Vol. 39, Netherlands, pp. 113-295.
\bibitem{dalziel60}Dalziel JAW, Welch AJE (1960) Acta Cryst 13:965
\bibitem{petrov11} Petrov D, Angelov B, Lovchinov V (2011) J Sol-Gel Sci Technol 58:636
\bibitem{holmes} Holmes LM, Van Uitert LG, Hecker RR, Hull GW (1972) Phys Rev B 5:138
\bibitem{holmes72} Holmes LM, Van Uitert LG (1972) Phys Rev B 5:147
\bibitem{schu} Schuchert H, H\"ufner S, Faulhaber R (1969) Z Physik 222:105
\bibitem {Cashion68} Cashion JD, Cooke AH, Thorp TL, Wells MR (1968) J Phys C 1:539
\bibitem {Bidaux68} Bidaux A, Meriel P (1968) J Phys 29:220
\bibitem{petrov12} Petrov D, Angelov B (2011) Acta Phys Pol A 122:737
\bibitem{mydosh93} Mydosh JA (1993) Spin glasses: an experimental introduction. Taylor and Francis, London
\bibitem{rmp} Binder K, Young AP (1986) Rev Mod Phys 58:801
\bibitem{blundell} Blundell S(2011) Magnetism in Condensed Matter. Oxford University Press, New York
\bibitem{perez98} P\'erez J, Garc\'{\i}a J, Blasco J, Stankiewicz J (1998) Phys Rev Lett 80:2401
\bibitem{nam00} Nam DH, Mathieu R, Nordblad P, Khiem NV, Phuc NX (2000) Phys Rev B 62:8989
\bibitem{motohashi05} Motohashi T, Caignaert V, Pralong V, Hervieu M, Maignan A, Raveau B (2005) Phys Rev B 71:214424
\bibitem{rietveld} Young RA (1993) The Rietveld Method. Oxford University Press Inc., New York
\bibitem{Tanaeva03} Tanaeva IA, Ikeda H, van Bokhoven LJA, Matsubara Y, de Waele ATAM (2003) Cryogenics 43:441
\bibitem{tari}Tari A (2003) The specific heat of mater at low temperatures. Imperial College Press, London
\bibitem{miyazaki05} Miyazaki Y, Wang Q, Yu Q-S, et al. (2005) Thermochimica Acta 431:133
\bibitem{dug} Dugdale JS, MacDonald DKC (1954) Phys Rev 96:57
\bibitem{cashion98} Cashion JD, Wells MR (1998) J Magn Magn Mater 177-181:781
\bibitem{kuzmin93} Kuz'min MD, Tishin AM (1993) J Appl Phys 73:4083
\bibitem{gruber82} Gruber JB, Chirico RD, Westrum Jr EF (1982) J Chem Phys 76:4600
\bibitem{Krynetski97} Krynetski\u{\i} IB, Matveev VM (1997) Phys Solid State 39:584 
\bibitem{binder86} Binder K, Young AP (1986) Rev Mod Phys 58:801
\bibitem{gunnarsson88} Gunnarsson K, Svedlindh P, Nordblad P, Lundgren L, Aruga H, Ito A (1988) Phys Rev Lett 61:754
\bibitem{malinowski11} Malinowski A, Bezusyy VL, Minikayev R, Dziawa P, Syryanyy Y, Sawicki M (2011) Phys Rev B 84:024409
\bibitem{imamura08} Imamura N, Karppinen M, Motohashi T, Yamauchi H (2008) Phys Rev B 77:024422
\bibitem{tang06} Tang Y-K, Sun Y, Cheng Z-H (2006) Phys Rev B 73:0124009
\bibitem{de06} De K, Thakur M, Manna A, Giri S (2006) J Appl Phys 99:013908
\end{document}